\documentclass[aps,prd,showpacs,preprintnumbers,nofootinbib,floatfix,floats,groupedaddress,twocolumn]{revtex4}

\usepackage{bm}
\usepackage{latexsym}
\usepackage{dcolumn}
\usepackage{amsmath,amsfonts,amssymb}
\usepackage{graphicx,epsfig}
\usepackage{color}
\usepackage[active]{srcltx}
\usepackage{subfig}

\def\nn{\nonumber}

\def\l{\left}
\def\r{\right}
\def\DM{\mathrm{d}}

\newcommand{\xdownarrow}[1]{%
  \ensuremath{%
    \left\downarrow\vbox to #1{}\right.\kern-\nulldelimiterspace
  }%
}

\newcommand{\xuparrow}[1]{%
  \ensuremath{%
    \left\uparrow\vbox to #1{}\right.\kern-\nulldelimiterspace
  }%
}

\reversemarginpar

\def \gm {{\overset{\star}{g}}}
\def \mK {{\overset{\star}{K}}}
\def \mnabla {{\overset{\star}{\nabla}}}
\def \mGamma {{\overset{\star}{\Gamma}}}
\def \mRs {{\overset{\star}{R}}}
\def \macc {{\overset{\star}{a}}}

\def \lp {L_0}

\def \mA {{\Omega^2}}
\def \mB {\mathcal B}
\def \mD {\Theta}

\makeatletter

\renewenvironment{widetext@grid}{%
  \par\ignorespaces
  \setbox\widetext@top\vbox{%
   \vskip15\p@
   \hb@xt@\hsize{%
    \leaders\hrule\hfil
    \vrule\@height6\p@
   }%
   \vskip6\p@
  }%
  \setbox\widetext@bot\hb@xt@\hsize{%
    \vrule\@depth6\p@
    \leaders\hrule\hfil
  }%
  \onecolumngrid
  \let\set@footnotewidth\set@footnotewidth@ii
}{%
  \par
  \twocolumngrid\global\@ignoretrue
  \@endpetrue
}%

\makeatother

\begin{document}

\title{Intrinsic and Extrinsic curvatures in Finsler{\it esque} spaces}

 \author{Dawood Kothawala}
 \email{dawood@physics.iitm.ac.in}
 \affiliation{Department of Physics, Indian Institute of Technology Madras, Chennai 600 036}

\date{\today}
\begin{abstract}
\noindent We consider metrics related to each other by functionals of a scalar field $\varphi(x)$ and it's gradient $\bm \nabla \varphi(x)$, and give transformations of some key geometric quantities associated with such metrics. Our analysis provides useful and elegant geometric insights into the roles of {\it conformal} and {\it non-conformal} metric deformations in terms of intrinsic and extrinsic geometry of $\varphi$-foliations. As a special case, we compare {\it conformal} and {\it disformal} transforms to highlight some non-trivial scaling differences. We also study the geometry of {\it equi-geodesic} surfaces formed by points $p$ at constant geodesic distance $\sigma(p,P)$ from a fixed point $P$, and apply our results to a specific disformal geometry based on $\sigma(p,P)$ which was recently shown to arise in the context of spacetime with a minimal length.
\end{abstract}

\maketitle
\vskip 0.5 in
\noindent
\maketitle
\section{Introduction} \label{sec:intro} 
Consider a spacetime described by a metric $\bm g$, and let $\varphi(x)$ be a given scalar field. From these variables, one can construct another metric on the same manifold defined by 
\begin{eqnarray}
\gm_{ab} &=& \mA g_{ab} - \epsilon \mB t_a t_b
\label{eq:qmetric}
\end{eqnarray}
where $\Omega=\Omega[\varphi]$ and $\mB[\varphi]$ are arbitrary functions of the scalar field $\varphi$, and
\begin{eqnarray}
t_a &=& \frac{\partial_a \varphi}{\sqrt{ \epsilon g^{ij} \nabla_i \varphi \nabla_j \varphi }} 
\;\; ; \;\;\;
g^{ab} t_a t_b = \epsilon = \pm 1
\end{eqnarray}

Our aim in this paper would be study the relationship between geometric quantities associated with $\bm \gm$ and $\bm g$.

However, before doing that, let us first give a brief motivation for considering such geometries. Under what circumstances can such geometries arise, and what might be their use in physical theories? Over the past several decades, it has been realized (although not widely appreciated) that the geometry relevant for describing matter in presence of gravitational field could be related to the spacetime geometry in a non-trivial manner, going beyond the well known relation based on {\it conformal transformations} which have otherwise been ubiquitous in such contexts. A very transparent presentation of this argument can be found, for example, in Bekenstein's paper \cite{bek-disformal}. He argued that if one takes {\it Finsler}, rather than {\it Riemannian}, geometry as more fundamental, and imposes some physically motivated constraints, a Finsler geometry can be re-cast in terms of a Riemannian geometry with a metric that is related to the spacetime metric by transformations which are non-conformal. Since the transformation in (\ref{eq:qmetric}) is qualitatively of such form (although differs in detail from the ones in \cite{bek-disformal}), we refer to such geometries as ``Finsler{\it esque}", and shall be working throughout this paper in this restricted 
context. A more recent reason for considering such geometries comes from semi-classical and quantum gravity. In particular, it was recently shown \cite{dk-ml} that space(time)s with a minimal length scale might be described by an {\it effective} ``metric" of the form (\ref{eq:qmetric}), with the non-local Synge world-function bi-scalar replacing the field $\varphi(x)$. We discuss this geometry in detail in Section \ref{sec:min-length} below.

{\it Summary}: I outline a method which gives, in a convenient form, the expressions for Ricci scalar associated with the metric $\gm_{ab}$ in terms of geometric quantities associated with the level surfaces $\Sigma$ of $\varphi(x)$. In Section \ref{sec:extrinsic}, I focus on the geometry of $\Sigma$ and find several relationships between intrinsic and extrinsic geometric quantities associated with $\Sigma$ as embedded in $\gm_{ab}$ and $g_{ab}$. In Section \ref{sec:modified-riccis}, I use these relations in the Gauss-Codazzi equation to re-construct the Ricci scalar of $\bm \gm$ (see Eqs.~(\ref{eq:mRs-final}, \ref{eq:JcJd})), and give several specific applications, including a recent result in which a non-local disformal coupling based on Synge world function arises naturally in spacetimes with a minimal length. In Section \ref{sec:riccis2}, I recast the derived expression for the Ricci scalar in a manner which elegantly highlights the (well-known) contribution of the conformal part, and the additional contribution due to the non-conformal part expressed in a transparent, geometric way. Finally, I end with brief concluding remarks in Section \ref{sec:concluding-remarks}.

{\it Notation}: The signature is $(-,+,+, \ldots)$ for Lorentzian spaces. I will also use the convenient notation $D_k=D-k$ which is handy when working in $D$ dimensions.
\section{Intrinsic and Extrinsic geometry} \label{sec:extrinsic} 
Given the metric $\gm_{ab}$, the inverse metric is easily found to be
\begin{eqnarray}
\gm {}^{ab} &=& \frac{1}{\mA} \; g^{ab} + \epsilon \l( \frac{\Omega^{-2} \mB}{\mA-\mB} \r) q^a q^b
\end{eqnarray}
where $q^a = g^{ab} t_b$ and $\gm {}^{ai} \gm_{ib} = \delta^a_{\phantom{a}b}$. 

To begin with, introduce the vectors 
\begin{eqnarray}
T_a &=& \sqrt{\mA-\mB} \; t_a
\nn \\
T^a &=& \gm {}^{ab} T_b
\nn \\
&=& \frac{1}{\sqrt{\mA-\mB}} \; q^a
\end{eqnarray}
which are normalized with respect to $\gm_{ab}$. Of course, any characterization of $\Sigma$ in $\gm_{ab}$ must be based on these vectors. The metric determinants are related by
\begin{eqnarray}
\sqrt{-\gm} = \Biggl\{ \Omega^{(D-2)} \sqrt{\frac{\mA-\mB}{\Omega^{-2}}} \Biggl\} \sqrt{-g}
\end{eqnarray}
which follows from the {\it matrix determinant lemma}: 
\begin{equation}
{\mathrm{det}} \l(\rm \bf M + \rm \bf u \rm \bf v^{\mathrm{T}}\r) = \l(\mathrm{det} \; \rm \bf M\r) \times \l( 1 + \rm \bf v^{\mathrm{T}} \rm \bf M^{-1} \rm \bf u \r)                                                                                                                                                                         \end{equation} 
where $\rm \bf M$ is an invertible square matrix, and $\rm \bf u, \rm \bf v$ are column vectors (of same dimension as $\rm \bf M$).

\begin{center}
\textbf{The first fundamental form}
\end{center}

Using the above relations, one can immediately deduce the following relation between the induced metrics, or the first fundamental forms, of $\Sigma$ in $\bm \gm$ and $\bm g$. 
\begin{eqnarray}
\overset{\star}{h}_{ab} &=& \gm_{ab} - \epsilon T_a T_b
\nn \\
&=& \l( \mA g_{ab} - \epsilon \mB t_a t_b \r) - \epsilon \l( \mA-\mB \r) t_a t_b
\nn \\
&=& \mA \; h_{ab}
\end{eqnarray}
That is, the {\it induced} geometries on $\Sigma$ in any two space(time)s related by Eq.~(\ref{eq:qmetric}) are related by a {\it conformal transformation}. 

This observation will enormously simplify the evaluation of Ricci scalar of $\gm_{ab}$ in terms of quantities associated with $g_{ab}$.

\begin{center}
\textbf{The intrinsic Ricci scalar}
\end{center}

The above result implies that intrinsic geometries of $\Sigma$ are conformal to each other, and since the conformal factor is $\mA[\varphi]$ which is constant on $\Sigma$, one immediately obtains
\begin{eqnarray}
\overset{\star}{R}_{\Sigma} = \Omega^{-2} R_{\Sigma}
\\
\nn
\end{eqnarray}
which is a simple rescaling, and, most importantly, independent of $\mB$.

\begin{center}
\textbf{The second fundamental form}
\end{center}

Our next aim would be relate the extrinsic geometries, or the second fundamental forms, of $\Sigma$ in the two metrics. These are defined by
\begin{eqnarray}
K_{ij} &=& \nabla_i t_j - \epsilon a_j t_i
\nn \\
\mK_{ij} &=& \mnabla_i T_j - \epsilon \macc_j T_i
\end{eqnarray}
where $a_j$ and $\macc_j$ are the acceleration vectors associated with vectors $q^i$ and $T^i$ respectively.
\begin{eqnarray}
a_j &=& q^k \nabla_k t_j
\nn \\
\macc_j &=& T^k \mnabla_k T_j
\end{eqnarray}

Since $\mnabla_b T_c = \partial_b T_c - \mGamma {}^a_{\phantom{a}bc} T_k$, we need the relation between Christoffel connections of the two metrics. This is given by 
\begin{eqnarray}
\mGamma {}^a_{\phantom{a}bc} = \Gamma {}^a_{\phantom{a}bc} + \frac{1}{2} \gm {}^{am} 
\l( - \nabla_m \gm_{bc} + 2 \nabla_{(b} \gm_{c)m} \r)
\label{eq:chr-rel}
\end{eqnarray}
We only need $\mGamma {}^k_{\phantom{k}ij} T_k$, which is relatively straightforward to obtain using the identities proved in Appendix \ref{app:chr}, from which we obtain
\begin{eqnarray}
\mnabla_b T_c &=& \sqrt{\mA-\mB} \; \l( \nabla_b t_c + \frac{\nabla_{\bm q} \mA}{2 \l(\mA-\mB\r)} h_{bc} + \frac{\mB}{\mA-\mB} K_{(bc)}  \r)
\nn
\end{eqnarray}
It immediately follows that
\begin{eqnarray}
\macc_c &=& T^b \mnabla_b T_c
\nn \\
&=& \frac{1}{\sqrt{\mA-\mB}} q^b \mnabla_b T_c
\nn \\
&=& a_c
\end{eqnarray}

Putting everything together, and noticing that hypersurface orthogonality implies $K_{(bc)}=K_{bc}$, we finally obtain
\begin{eqnarray}
\mK_{ab} &=& \mnabla_{a} T_b - \epsilon \macc_b T_a
\nn \\
&=& \frac{1}{\sqrt{\mA-\mB}} \Biggl[ \mA K_{ab} + \frac{1}{2} \l( \nabla_{\bm q} \mA \r) h_{ab} \Biggl]
\end{eqnarray}
and
\begin{eqnarray}
\overset{\star}{K} &=& \gm {}^{ab} \mK_{ab} 
\nn \\
&=& \frac{\Omega^{-2}}{\sqrt{\mA-\mB}} \Biggl[ \mA K + \frac{D_1}{2} \nabla_{\bm q} \mA \Biggl]
\end{eqnarray}

\begin{center}
\textbf{Summary}
\end{center}

To summarize, we have derived the following relations between first and second fundamental forms of $\varphi(x)=$constant
\begin{eqnarray}
\overset{\star}{h}_{ab} &=& \mA h_{ab}
\nn \\
\overset{\star}{R}_{\Sigma} &=& \Omega^{-2} R_{\Sigma}
\nn \\
\mK_{ab} &=& \frac{\mA}{\sqrt{\mA-\mB}} \Biggl[ K_{ab} + \l( \nabla_{\bm q} \ln \Omega \r) h_{ab} \Biggl]
\nn \\
\mK &=&  \frac{1}{\sqrt{\mA-\mB}} \Biggl[ K + D_1 \nabla_{\bm q} \ln \Omega \Biggl]
\end{eqnarray}

Of special interest are two cases which are displayed in Table~\ref{tab:conformal-vs-disformal}, corresponding to conformal and disformal transformations.

\section{Re-constructing the Ricci scalar from geometry of $\Sigma$} \label{sec:modified-riccis}

Having the relationship between intrinsic and extrinsic geometrical properties of $\Sigma$, we can now re-construct the point wise Ricci scalar of $\gm_{ab}$ by using the Gauss-Codazzi relation:
\begin{eqnarray}
\mRs = \overset{\star}{R}_{\Sigma} - \epsilon \l( \mK {}^2 + \mK {}_{ab}^2 \r) - 2 \epsilon \mnabla_{\bm T} \mK + 2 \epsilon \mnabla_i \macc^i
\label{eq:modrs1}
\end{eqnarray}
where $\mnabla_{\bm T} \equiv T^i \mnabla_i$ and $\mK {}_{ab}^2 \equiv \gm {}^{ia} \gm {}^{jb} \mK_{ab} \mK_{ij}$. 

Our aim is to express all the quantities on the RHS of the above equation in terms of quantities associated with $g_{ab}$, thereby obtaining the transformation of $\mRs$. This is easily done by using the results of the previous sections, which give the required relationships for quantities appearing in RHS above. 

At this stage, it is useful to trade off the function $\mB$ in terms of a new function $\alpha$, defined by 
\begin{eqnarray}
\mA - \mB = \alpha^{-1}
\label{eq:alphadef}
\end{eqnarray}
The significance of introducing $\alpha$ will become clear as we go along. In particular, the set of metric transformations of the form (\ref{eq:qmetric}) turn to have an extremely 
simple composition law in terms of the functions $\l(\mA, \alpha\r)$; see Appendix \ref{app:composition-law}.

Using expressions from previous section, one can derive the following relations after few (long) algebraic steps:
\begin{eqnarray}
\mK {}_{ab}^2 &=& \alpha \l[ K_{ab}^2 + K \nabla_{\bm q} \ln \mA + D_1^2 \l(  \nabla_{\bm q} \ln \Omega \r)^2 \r]
\nn \\
\mK {}^2 &=& \alpha \l[ K^2 + D_1 K \nabla_{\bm q} \ln \mA + D_1^2 \l(  \nabla_{\bm q} \ln \Omega \r)^2 \r]
\nn \\
\mnabla_{\bm T} \mK &=& \alpha \nabla_{\bm q} K + \frac{1}{2} \l( K +  D_1 \nabla_{\bm q} \ln \Omega \r) \nabla_{\bm q} \alpha
\nn \\
\phantom{\mnabla_{\bm T} \mK} && \phantom{\alpha \nabla_{\bm q} K } + D_1 \alpha \nabla_{\bm q} \nabla_{\bm q}  \ln \Omega
\nn \\
\mnabla_i \macc^i &=& \Omega^{-2} \nabla_i a^i
\label{eq:aux-eqs}
\end{eqnarray}
where $\nabla_{\bm q} \equiv q^i \nabla_i$ as above. 

We now have everything needed to evaluate the RHS of Eq.~(\ref{eq:modrs1}). Putting everything together, and using the (easily proved) identity 
\begin{eqnarray}
\nabla_{\bm q} K &=& q^i \nabla_i \nabla_j q^j
\nn \\
&=& - K_{ab}^2 - R_{ab} q^a q^b + \nabla_i a^i
\end{eqnarray}
we get
\begin{eqnarray}
\mRs = \Omega^{-2} R
\;+\; \epsilon \l( \alpha - \Omega^{-2} \r)  \mathcal{J}_d 
\;-\; \epsilon \alpha \, \mathcal{J}_c %
\label{eq:mRs-final}
\end{eqnarray}  
where
\begin{eqnarray}
\mathcal{J}_c &=&
\epsilon \l[ 2 D_1 \Omega^{-1} \square \Omega + D_1 D_4 \Omega^{-2} (\nabla \Omega)^2 \r]
\nn 
\\
\vspace{.2cm}
&& \hspace{.55cm} \; + \l( K + D_1 \nabla_{\bm q} \ln \Omega \r) \times \nabla_{\bm q} \ln \alpha \Omega^2 
\nn \\
\nn \\
\mathcal{J}_d &=&
2 R_{ab} q^a q^b + K_{ab}^2 -K^2 - 2 \nabla_i a^i 
\nn \\
&=& \epsilon \l( R - R_{\Sigma} - 2 \nabla_i a^i \r)
\label{eq:JcJd}
\end{eqnarray}

which is the required expression. 

The crucial role played by the {\it non-conformal} term in the metric Eq.~(\ref{eq:qmetric}), characterized by $\mB \neq 0$, is immediately obvious from the above relation. In particular, $ \mathcal{J}_d$ in the above expression is multiplied by $\alpha-\Omega^{-2} = \alpha \Omega^{-2} \mB$ (see Eq.~(\ref{eq:alphadef})). For conformally related metrics, $\mB=0$, and therefore this term does not play any role in conformal transformations. This observation is crucial, since it couples the non-conformal part of the metric $\gm_{ab}$ to the object $\mathcal{J}_d$ which has a very special structure!
\\
\subsection{Applications and Analyses}

\subsubsection{Raychaudhuri equation}

The first application we give is based on the third relation in Eqs.~(\ref{eq:aux-eqs}). 

Consider the congruence of integral curves of $q^i$, which need not necessarily be geodesics. Then, the expansion associated with these congruences is $\theta=K$, and it's rate of change along the curves given by $\dot \theta=\nabla_{\bm q} K$. The corresponding quantities in metric $\gm_{ab}$ are given by $\overset{\star}{\theta}=\mK$ and $\overset{\star}{\dot \theta}=\mnabla_{\bm T} \mK $, and their relationship, translated to variables more familiar from the Raychaudhuri equation, is given by
\begin{eqnarray}
\overset{\star}{\l(\frac{\DM \theta}{\DM \lambda}\r) } &=& \alpha \frac{\DM \theta}{\DM \lambda} +  \frac{1}{2} \l( \theta +  D_1 \frac{\DM \ln \Omega}{\DM \lambda} \r) \frac{\DM \alpha}{\DM \lambda} + D_1 \alpha \frac{\DM^2 \ln \Omega}{\DM \lambda^2}
\nn \\
\end{eqnarray}
where we have replaced $\nabla_{\bm q}$ on the RHS with $\DM/\DM \lambda$ - the derivative along the curve. (The special case of conformal transformation corresponds to $\alpha=1/\mA$.) 

The above relation gives the modification to the Raychaudhuri equation, and hence should be useful in studying the focussing and de-focussing behavior of geodesics in Finsler-Riemann spaces. If the origin of the scalar field $\varphi(x)$ is due to some fundamental physics associated with quantum gravitational effects, such a modification would be key to understand the effect of quantum gravity on spacetime singularities.

In this context, it is also worth pointing out the relation between the symmetric, traceless part of $K_{ab}$, or the so-called shear-tensor, which happens to be much simpler that the relation between $K_{ab}$ themselves. Using again Eqs.~(\ref{eq:aux-eqs}), we get
\begin{eqnarray}
\overset{\star}{\sigma}_{ab} &=& \mK_{ab} - (1/D_1) \mK \; \overset{\star}{h}_{ab}
\nn \\
&=& \sqrt{\alpha} \; \mA \sigma_{ab}
\end{eqnarray}
and hence
\begin{eqnarray}
\overset{\star}{\sigma} {}^2_{ab} &=& \alpha \; \sigma^2_{ab}
\end{eqnarray}
It is particularly obvious from the above that the conformal ($\alpha=1/\mA$) and disformal ($\alpha=\mA$) cases correspond to completely different scaling of the shear tensor associated with $\varphi(x)$ foliations.

A more general discussion on the propagation of expansion, shear and vorticity in the context of Finsler-Riemann geometries can be found, for example, in \cite{FR-refs}. 

\subsubsection{Conformal transformation}

As mentioned at the end of previous section, for conformal transformations, $\alpha=1/\mA$, and the expression for Ricci scalar reduces to 
\begin{eqnarray}
\mRs &=& \Omega^{-2} R - \epsilon \Omega^{-2} \l[\mathcal{J}_c\r]_{\alpha \Omega^2=1}
\nn \\
\nn \\
&=& \Omega^{-2} \l[ R - 2 D_1 \Omega^{-1} \square \Omega - D_1 D_4 \Omega^{-2} (\nabla \Omega)^2 \r]
\end{eqnarray}
since $\ln \alpha \Omega^2 = 0$ in this case, and hence the second term in $\mathcal{J}_c$ in Eq.~(\ref{eq:JcJd}) vanishes. 

We have therefore re-derived the well known expression for conformal transformation of the Ricci scalar, in a manner which gives a geometric origin for the $\Omega$ dependent  terms.

\subsubsection{Disformal transformation}

For the special case of disformal transformations
\footnote{We should clarify that we are using the term {\it disformal} here to refer to a special subclass of metrics (\ref{eq:qmetric}), while sometimes all 
such metrics 
are called disformal. This is just a matter of terminology.}
, $\alpha=\Omega^2$, and we obtain
\begin{eqnarray}
\mRs = \Omega^{-2} R + \epsilon \l( \Omega^2 - \Omega^{-2} \r)  \mathcal{J}_d - \epsilon \Omega^2 \l[\mathcal{J}_c\r]_{\alpha = \Omega^2}
\label{eq:Rsdisformal}
\end{eqnarray}

Since disformal modifications of spacetime geometry play an important role in several studies such as modified gravity, the above expression can provide considerable insight into construction of sensible action for such models.

\subsubsection{Small scale structure of spacetime: \\ Disformal coupling through Synge world function bi-scalar} \label{sec:min-length}

\begin{figure*}%
    \centering
    \subfloat[The geometry of {\it equi-geodesic} ``foliation".]{{\includegraphics[width=0.7\textwidth]{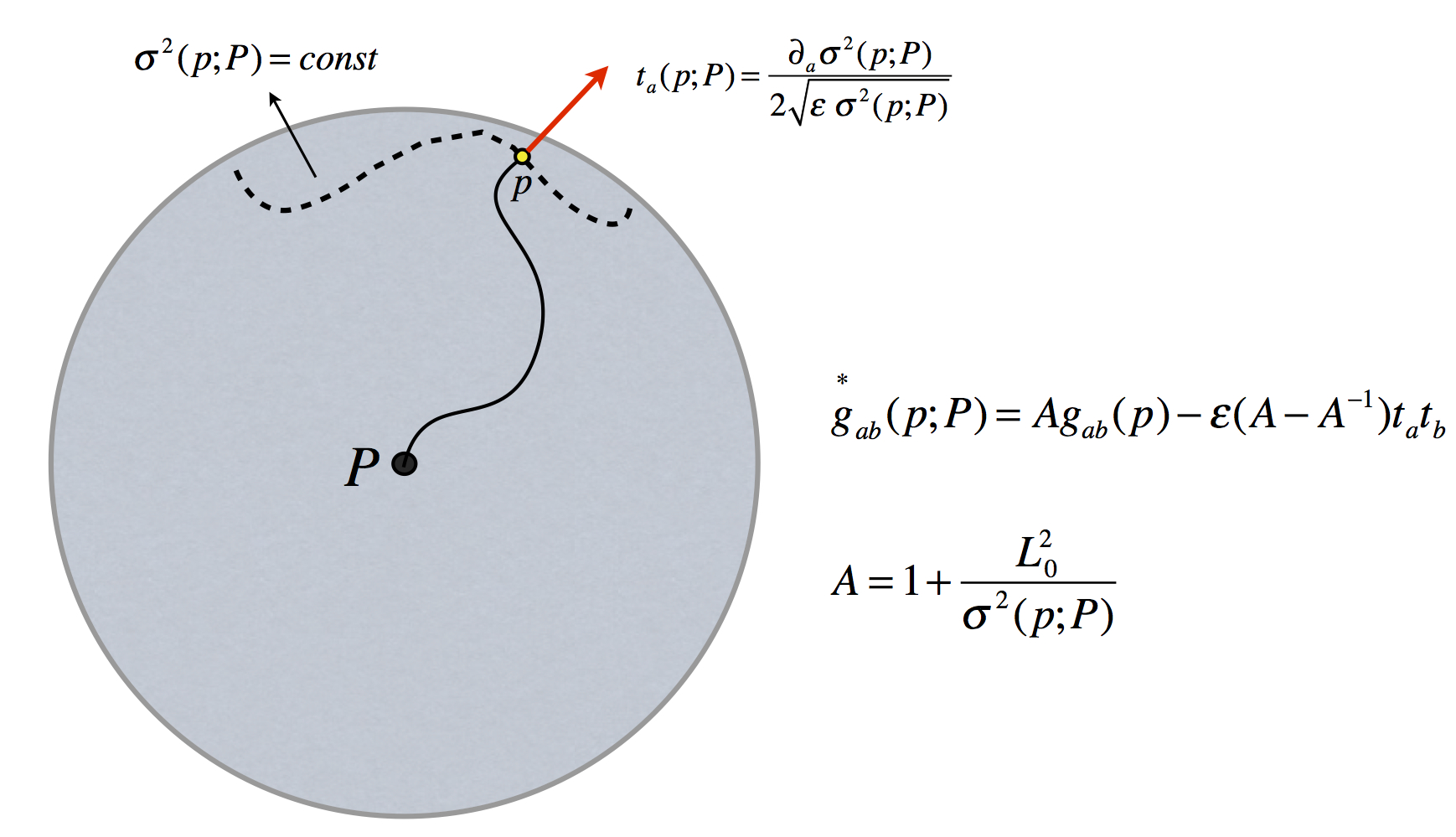} }}%
    \subfloat[Equi-geodesic surface $\Sigma$ in Minkowski spacetime. $\mathcal{H}$ is the null cone at $P$.]{{\includegraphics[width=0.325\textwidth]{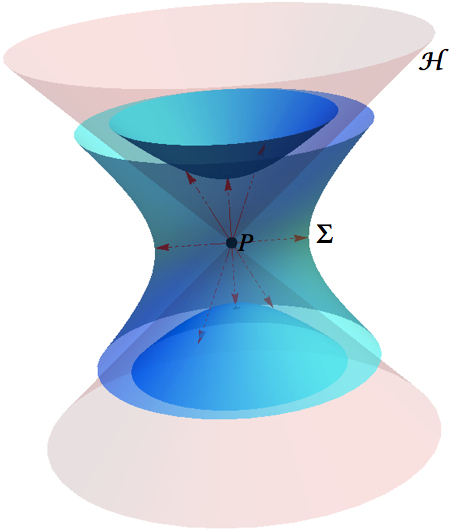} }}%
    \caption{Set of events $p$ at constant geodesic distance from a given space(time) event $P$. The shaded region represents the normal convex neighborhood $\mathcal N(P)$ 
	of $P$, and $p \in \mathcal N(P)$.}%
    \label{fig:geodesic-congruence}%
\end{figure*}

In a recent work \cite{dk-ml}, it was argued that a spacetime with a minimal length is endowed, under certain conditions, with precisely a disformal structure based on the bi-scalar of geodesic distance, $\sigma(p,P)$, between spacetime events $p$ and $P$, with
\begin{eqnarray}
\mA = 1 + \frac{\lp^2}{\sigma(p,P)^2}
\end{eqnarray}
It is a straight forward exercise to plug the above form for $\Omega$ in Eq.~(\ref{eq:Rsdisformal}) and obtain Eq.~(14) of \cite{dk-tp}. 

{\it Equi-geodesic surfaces}: 

The key role here is played by the congruence of geodesics emanating from a fixed spacetime event $P$, and the surface comprised of events $p$ lying at constant geodesic interval from $P$, which we call as the {\it equi-geodesic} surface. We give below several properties of the foliation based on such surfaces, and also sketch the derivation of Eq.~(14) in \cite{dk-tp} which was based on the analysis presented here. Let us start with the key geometric quantities associated with $\sigma^2=$constant surfaces with one of the events, say $P$, is fixed; see Fig.\;\ref{fig:geodesic-congruence}. 

Since the affinely parametrized tangent vector to the geodesic connecting $P$ to $p$ is the normal to the $\Sigma$, and given by  \cite{poisson-lrr}
\begin{eqnarray}
t_a = \frac{\nabla_a \sigma^2}{2 \sqrt{\epsilon \sigma^2}}
\end{eqnarray}
the extrinsic curvature tensor of $\sigma^2=$constant surface, $\Sigma$, is given by
\begin{eqnarray}
K_{ab} &=& \nabla_a t_b
= \frac{\nabla_a \nabla_b \l( \sigma^2/2 \r) - \epsilon t_a t_b}{\sqrt{\epsilon \sigma^2}}
\end{eqnarray}
This particular foliation, formed out of points which are at a fixed geodesic interval from a given point, has many interesting characteristics, all deriving from the fact that the bi-tensor 
$\nabla_a \nabla_b \l( \sigma^2/2 \r) $ has a well know covariant Taylor series expansion at $p$ near $P$ \cite{christenson-worldf}:
\begin{eqnarray}
\nabla_a \nabla_b \l( \frac{1}{2} \sigma^2 \r) &=& g_{ab} - \frac{\lambda^2 }{3} \mathcal{E}_{ab} + \frac{\lambda^3}{12} \nabla_{\bm q} \mathcal{E}_{ab} 
\nn \\
&-& \frac{\lambda^4}{60} \l(\nabla^2_{\bm q} \mathcal{E}_{ab} + \frac{4}{3} \mathcal{E}_{ia} \mathcal{E}^i_{\phantom{i}b} \r) + O(\lambda^5)
\nn \\
\end{eqnarray}
where $\nabla_{\bm q} \equiv q^i \nabla_i$, $\mathcal E_{ab} = R_{a m b n} q^m q^n$, and $\lambda=\sqrt{\epsilon \sigma^2}$ is the numerical value of the geodesic distance between $P$ and $p$. 

Therefore, we see that the extrinsic geometry of such a equi-geodesic ``foliation" is very special, and completely characterized by the {\it tidal tensor} $\mathcal E_{ab} = R_{a m b n} q^m q^n$. In fact, the intrinsic and extrinsic curvatures can be characterized by a systematic Taylor expansion around $P$, given by
\begin{eqnarray}
K_{ab} &=& \frac{1}{\lambda}h_{ab} - \frac {1} {3} \lambda \mathcal{E}_{ab} + \frac {1} {12} \lambda^2 \nabla_{\bm q} \mathcal{E}_{ab} - \frac {1} {60}\lambda^3 F_{ab} + O(\lambda^4)
\nn \\
\nn \\
K &=& \frac{D_1}{\lambda} - \frac {1} {3} \lambda \mathcal{E} + \frac {1} {12} \lambda^2 \nabla_{\bm q} \mathcal{E} - \frac {1} {60}\lambda^3 F + O(\lambda^4)
\nn \\
\nn \\
R_{\Sigma} 
&=& \frac {\epsilon D_1 D_2} {\lambda^2} + R - \frac{2\epsilon(D+1)}{3} \mathcal{E} + O(\lambda)
\end{eqnarray}
where $\mathcal{E}=g^{ab} \mathcal{E}_{ab} = R_{ab} q^a q^b$, $F_{ab} = \nabla_{\bm q}^2 \mathcal{E}_{ab} + (4/3) \mathcal{E}_{ak} \mathcal{E}^k_{\phantom{k}b}$, and $F = F_{ab} g^{ab}$.

The {\it exact} form of the Ricci scalar (in which the above Taylor expansions can be plugged if needed) turns out to be 
\begin{eqnarray}
\mRs_P(p) &=& \mA \; R(p) - \l( \mA - \Omega^{-2} \r) \times \l( R_{\Sigma} -  R^{\rm flat}_{\Sigma} \r)
\nn \\
&& \hspace{.1cm} + \;
2 \epsilon (\mA-1) \l(D_{-1}/D_1\r) K^{\rm flat} \times \l( K - K^{\rm flat} \r) 
\nn \\
\end{eqnarray}
where $R^{\rm flat}_{\Sigma}={D_1 D_2}/{\sigma^2}$ and $K^{\rm flat}={D_1}/{\sqrt{\epsilon \sigma^2}}$ are the induced and extrinsic curvatures of $\sigma^2=$const. surfaces in {\it flat} spacetime. Note that, in flat space(time), these surfaces are simply hyperboloids, and hence maximally symmetric $(D-1)$ spaces with positive or negative curvature; see Fig.\;\ref{fig:geodesic-congruence}(b).

Let us point out some special features of the above expression, which are absent for a generic $\varphi(x)$, being a consequence of (i) disformal nature of the coupling, (ii) the very specific form of $\mA=1+\lp^2/\sigma^2$ which implies that $\Sigma$ corresponds to $\sigma^2=$ constant surfaces.

1. The first two terms on RHS mimic the relationship between $g_{ab}$ and $\gm_{ab}$. 

2. The form of the RHS clearly indicates an interplay between $\mA=1$ and $\bm g \equiv$ Riemann flat: the disformal character of the modification $\mA \neq 1$ only couples to the curvature of the background spacetime, that is: {\it flat space(time) is immune to the disformal modification of the above form}!

3. The geometrical structure of the above expression might hold the key to a generic study of behavior a disformal spacetime near curvature singularities of $\bm g$ in terms of focusing and de-focussing of geodesics. 

\begin{widetext}
\newsavebox\Conf
\begin{lrbox}{\Conf}
  \begin{minipage}{0.5\textwidth}
    \begin{align*}
     \gm_{ab} &= \mA g_{ab} \\
     & \\
     \overset{\star}{h}_{ab} &= \mA h_{ab} \\
     & \\
      \mK_{ab} &= \Omega K_{ab} + \l( \nabla_{\bm q} \Omega \r) h_{ab} \\
      & \\
      \mK &= \Omega^{-1} K + D_1 \Omega^{-2} \nabla_{\bm q} \Omega 
    \end{align*} 
  \end{minipage}
\end{lrbox}

\newsavebox\Disf
\begin{lrbox}{\Disf}
  \begin{minipage}{0.5\textwidth}
    \begin{align*}
    	\gm_{ab} &= \mA g_{ab} - \epsilon \l( \mA - \Omega^{-2} \r) t_a t_b \\
	& \\
          \overset{\star}{h}_{ab} &= \mA h_{ab} \\
          & \\
          \mK_{ab} &= \Omega^3 K_{ab} + \l( \mA \nabla_{\bm q} \Omega \r) h_{ab} \\
          & \\
      \mK &= \Omega K + D_1 \nabla_{\bm q} \Omega 
    \end{align*} 
  \end{minipage}
\end{lrbox}

\begin{table}[h]
  \begin{center}
    \begin{tabular}{ccc}
    \\ \hline \hline \\
      Conformal transformations & & Disformal transformations \\ 
      & & \\
      $\mB=0; \;\; \alpha = \Omega^{-2}$ & & $\mB = \mA - \Omega^{-2}; \;\; \alpha=\mA$ \\ \hline \hline
      \usebox{\Conf} & & \usebox{\Disf} \\
        \\ \hline \hline \\
    \end{tabular}
  \end{center}
\caption{Comparison of conformal and disformal transforms of {\it first} and {\it second} fundamental forms.}
  \label{tab:conformal-vs-disformal}
\end{table}
\vspace{1cm}
\end{widetext}

\section{Conformal and non-conformal parts of the  Ricci scalar} \label{sec:riccis2}

Let ${\bf Ric}\l[\bm g\r]$ denote the Ricci scalar associated with metric $\bm g$, and define $\mD = \alpha \Omega^2 - 1$ which measures deviation from the conformal case, for which $\alpha = \Omega^{-2}$ and hence $\mD=0$. Then, recalling that
	\begin{eqnarray}
	R_{\Sigma, \bm g} = {\bf Ric}\l[ \bm g \r] + \epsilon \l( K_{ab}^2 + K^2 \r) + 2 \epsilon \nabla_{\bm q} K - 2 \epsilon \bm \nabla \bm \cdot \bm a
	\nn
	\end{eqnarray}		
	is the intrinsic Ricci scalar of the $\Sigma$ foliation embedded in $\bm g$, Eq.~(\ref{eq:mRs-final}) can be cast in the following elegant form:
	\begin{widetext}
	\begin{eqnarray}
	{\bf Ric}\l[\mA \bm g - \epsilon \alpha^{-1} \mD \; \bm t \otimes \bm t\r] \; &=& \;
	 \underbrace{
	 \l( 1 + \mD \r) \; {\bf Ric}\l[\mA \bm g \r]
	 }_{\rm the~conformal~part} 
	- \; \;
	\Omega^{-2}
	 \underbrace{\l[\;
	 	 \mD \; \overbrace{\l( R_{\Sigma, \bm g} + 2 \epsilon \bm \nabla \bm \cdot \bm a \r)}^{\rm purely~{\it intrinsic}} 
		- \;
		\epsilon \dot \mD \; 
		\overbrace{\l( K + D_1 \nabla_{\bm q} \ln \Omega \r)}^{\rm purely~{\it extrinsic}} 
	\;\r]}_{{\rm contribution~of~the~} \bm t \otimes \bm t {\rm ~term}
	}
	\nn \\
	\nn \\
	\nn \\
		\; &=& \;
 	 \l( 1 + \mD \r) \; {\bf Ric}\l[\mA \bm g \r]
	- \; \;
	 	 \mD \; \l( R_{\Sigma} + 2 \epsilon \bm \nabla \bm \cdot \bm a \r)_{{\mA \bm h}}
		+ \;
		\epsilon \dot \mD \; 
		\Omega^{-1}
			K_{\Sigma, {\mA \bm h}}
	\label{eq:rs-geom-struc}
	\end{eqnarray}
	\end{widetext}	
where $\dot \mD = \nabla_{\bm q} \mD$.

We believe that the above expression holds the key to the importance of the non-conformal term in Finsler like spaces. It brings in some important geometric quantities associated with the $\varphi$ foliation, and drastically alters the behavior of the Ricci scalar in a manner which pure conformal transformations can not. 

The above fact was drastically brought to focus recently in the context of small scale structure of spacetime in presence of a minimal length \cite{dk-tp}, where the non-conformal part leads to some remarkable cancellations which drastically alters the low energy manifestation of minimal length effects, leaving a finite, $O(1)$, relic. This suggests that the non-conformal transformations, if they arise due to quantum gravitational effects, might have non-trivial effect on physics at all scales.

Besides the above, the geometric approach advocated here should also find immediate application in several other physical contexts; we here mention a few, and also point out related generalizations.

{\bf Relativistic MOND}: The class of metrics we have considered here have been most prominent in the context of Modified Newtonian Dynamics (MOND), where disformal transformations have played an 
	important role. The relativistic generalization of MOND, proposed by Bekenstein \cite{bek-mond}, involves a metric $\tilde g_{ab}$ constructed from the background metric $g_{ab}$, a 
	normalised timelike vector field $U_{a}$, and a scalar field $\varphi$. The structure of this modification has been extensively studied, specifically in the context of {\it cosmology}. 
	The geometric formalism presented here should 
	be useful to investigate the characteristics of $\tilde g_{ab}$, in particular, for better understanding of the contribution of non-conformal part of the deformation as well structure of modified action for the theory. 
	This is already suggested by the discussion in Section 
	\ref{sec:riccis2} above. Doing so would require a generalization of the present analysis to the case when the vector field in question $U_a \neq \nabla_a \varphi$. Such a generalization should be straightforward, and 
	we hope to address it in future work. We here simply illustrate the power of our method in a simplified context in cosmology.
	
	Consider the simplest (yet most studied) case of $k=0$ (flat), $D=4$, FLRW metric $g_{ab}$ in standard co-moving coordinates $(t,x,y,z)$ with scale factor $a(t)$. 
	To study deformations which respect the 
	background symmetry, we assume that $\Omega$ and $\mB$ are functions of $t$ only. Further, since conformal deformations are already 
	well understood, we set $\Omega=1$. In such a case, $\alpha=(1-\mB)^{-1}$, and the 
	relevant foliation is the one provided by $t=$constant surfaces. For such a foliation, $K=3 \l( \dot a/a \r)$ and $K_{ab}=(1/3) K h_{ab}$. Referring to Eq. (\ref{eq:rs-geom-struc}), and noting that $R_{\Sigma}=0=\bm a$, 
	we immediately get the modified Ricci scalar to be 
	\begin{eqnarray}
	{\bf Ric}\l[ g_{ab} + \mB \; \delta^0_a \delta^0_b \r] \; &=& \; \frac{ {\bf Ric}\l[ g_{ab} \r]}{1-\mB} - \frac{3 H(t) \dot \mB}{(1-\mB)^2}
	\end{eqnarray}
	where $H(t)=\dot a/a$ is the Hubble parameter. The non-flat case, $k = \pm 1$, is also easy to obtain by noting that the metric $h_{ab}$ (being maximally symmetric) immediately yields $R_{\Sigma, \bm h}=6k/a^2$, which 
	can then be plugged in Eq. (\ref{eq:rs-geom-struc}) along with other terms.
	
	{\bf Lanczos-Lovelock (LL) lagrangians}: 
	These class of lagrangians have played an important role in generalizing and understanding various classical and semi-classical aspects of gravity over the past decade 
	\cite{llreview}. Therefore, generalizing the calculation for Ricci scalar $R$ presented here to generic LL lagrangians would provide another most natural extension of the analysis presented here. Not only would this extend 
	the study of disformal modifications of gravity to Lanczos-Lovelock models of gravity, but it would open up a completely new avenue of research into semi-classical aspects of conformal and non-conformal deformations of 
	LL lagrangians. Let us briefly elaborate on this point. 
	The LL lagrangians $\mathcal{L}^{\rm bulk}_m$ of order $m$ are built from $m$ copies of curvature tensor $R_{abcd}$, while the surface term $\mathcal{K}^{\rm surface}_m$ 
	which makes the variational problem well defined is 
	built from sum of products of $^{(D-1)}R_{abcd}$ and $K_{ab}$. Therefore, although finding the bulk term requires the knowledge of full modified Riemann tensor (which remains a formidable task), the surface term $
	\mathcal{K}^{\rm surface}_m$, since it 
	involves only induced geometry $h_{ab}$, can be evaluated in a straightforward manner from the expressions given here. Since the surface term has played a very important role in semi-classical gravity, especially black 
	hole thermodynamics, it's 	disformal transformation can lead to new insights beyond those obtained from purely conformal transformations. We hope to take up this calculation in future work.
	

\section{Concluding Remarks} \label{sec:concluding-remarks}

The mathematical expressions derived here, and their geometric properties, should be useful in several contexts in which an object like $\gm_{ab}$ arises, either as a `physical metric' to which matter fields couple \cite{bek-disformal}, or through some effective quantum gravitational model \cite{dk-ml, dk-tp}.
We hope that these expressions and the method of derivation would be applicable to a wider class of problems in which the relevant scalar (or bi-scalar) degree of freedom arises through some physical consideration, and couples non-conformally to the spacetime geometry. Moreover, since the Ricci scalar forms the basis for setting up the standard gravitational action, the expression for $\mRs$ in the form given in Eq.~(\ref{eq:rs-geom-struc}) might provide insight into construction of the action involving $g_{ab}$ and $\varphi(x)$.

{\it Acknowledgements --} The author thanks IUCAA, Pune, where part of this work was done, for kind hospitality.

\appendix 
\section{Derivation of Christoffel connection component} 
\label{app:chr}

We start with the following relations which are straightforward to establish:
\begin{eqnarray}
q^m \nabla_m \gm_{bc} &=& \l( \nabla_{\bm q} \mA \r) g_{bc} - \epsilon \mB \l( t_b a_c + t_c a_b \r) - \epsilon \l(\nabla_{\bm q} \mB\r) t_b t_c
\nn \\
q^m \nabla_b \gm_{mc} &=& - \mB \nabla_b t_c + \epsilon \l[\nabla_{\bm q} \l(\mA-\mB\r)\r] t_b t_c
\label{eq:tmp1}
\end{eqnarray}
where $\nabla_{\bm q} \equiv q^m \nabla_m$. Using Eq.~(\ref{eq:tmp1}) in Eq.~(\ref{eq:chr-rel}), a few steps of algebra give
\begin{eqnarray}
\mGamma {}^a_{\phantom{a}bc} T_a &=& \sqrt{\mA-\mB} ~ \Gamma^a_{\phantom{a}bc} t_a + \frac{1}{2 \sqrt{\mA-\mB}} ~ \Delta_{bc}
\end{eqnarray}
where
\begin{eqnarray}
\Delta_{bc} &=& - \l( \nabla_{\bm q} \mA \r)  \; h_{bc} + \epsilon \l[\nabla_{\bm q} \l(\mA-\mB\r)\r] t_b t_c - 2 \mB K_{(bc)}
\nn
\end{eqnarray}
\\
\section{Composition law of transformations}
\label{app:composition-law}

As an interesting aside, we note the following composition law for the transformations we are considering in this paper. Using the variables $\l(A=\mA, \alpha\r)$, let us consider the following class of metrics, all defined on the same manifold.
\begin{eqnarray}
g^{(1)}_{ab} &=& A_{10} g^{(0)}_{ab} - \epsilon \l( A_{10} - \alpha^{-1}_{10} \r) t^{(0)}_a t^{(0)}_b
\nn \\
g^{(2)}_{ab} &=& A_{21} g^{(1)}_{ab} - \epsilon \l( A_{21} - \alpha^{-1}_{21} \r) t^{(1)}_a t^{(1)}_b
\end{eqnarray}
\\
Then, noting that $t^{(1)}_a = t^{(0)}_a/\sqrt{\alpha_{10}}$, it is easy to show that
\begin{eqnarray}
g^{(2)}_{ab} &=& \l(A_{21} A_{10}\r) g^{(0)}_{ab} - \epsilon \l[ A_{21} A_{10} - \l( \alpha_{21} \alpha_{10} \r)^{-1} \r] t^{(0)}_a t^{(0)}_b
\nn \\
\end{eqnarray}
which immediately yields the following simple composition law
\begin{eqnarray}
A_{20} &=& A_{21} A_{10}
\nn \\
\alpha_{20} &=& \alpha_{21} \alpha_{10} 
\end{eqnarray}
or symbolically
\begin{eqnarray}
{g^{(0)}_{ab}}
\;\; \overset{A_{10}, \alpha_{10}}{\xrightarrow{\hspace*{1cm}}} \;\;
g^{(1)}_{ab} 
\;\; \overset{A_{21}, \alpha_{21}}{\xrightarrow{\hspace*{1cm}}} \;\;
g^{(2)}_{ab}
\nn \\
\nn \\
{g^{(0)}_{ab}}
\;\; 
\underset{\alpha_{20}=\alpha_{21} \alpha_{10}}{
\overset{A_{20}=A_{21} A_{10}}{\xrightarrow{\hspace*{3.4cm}}} 
}
\;\;
g^{(2)}_{ab}
\end{eqnarray}


\widetext


\end{document}